# Topological insulator quantum dot with tunable barriers

Sungjae Cho[1], Dohun Kim, Paul Syers, Nicholas P. Butch[2], Johnpierre Paglione, and Michael S. Fuhrer*

*Center for Nanophysics and Advanced Materials, University of Maryland, College Park, MD 20742-4111, USA*

*to who correspondence should be address: mfuhrer@umd.edu.  [1]present address: Department of Physics and Materials Research Laboratory, University of Illinois at Urbana-Champaign, Urbana, Illinois 61801-2902, USA.  [2]present address: Condensed Matter and Materials Division, Lawrence Livermore National Laboratory, Livermore, CA 94550, USA.

**Thin (6-7 quintuple layer) topological insulator $Bi_2Se_3$ quantum dot devices are demonstrated using ultrathin (2~4 quintuple layer) $Bi_2Se_3$ regions to realize semiconducting barriers which may be tuned from Ohmic to tunneling conduction via gate voltage. Transport spectroscopy shows Coulomb blockade with large charging energy >5 meV, with additional features implying excited states.**

The three dimensional strong topological insulators (STIs) exhibit a bulk band gap and gapless Dirac surface states surface states on all surfaces.  The Dirac surface states are singly-degenerate, topologically protected from backscattering by time-reversal symmetry, and show spin-momentum locking. The STI $Bi_2Se_3$ has been studied by angle-resolved photoemission spectroscopy (ARPES)[1-3], scanning tunneling spectroscopy (STS)[4-6] and electrical transport measurements[7-10].  A significant challenge in the field of topological insulators is the design of mesoscopic devices (e.g. quantum dots, quantum point contacts) which are promising in both fundamental research on confined topological modes[11-14] as well as for spintronics[15] and quantum information applications[16-19] owing to the novel electronic structure of surface states. Dirac electrons cannot be confined by potentials due to Klein tunneling[20], hence gate confinement on the STI surface is impossible. Magnetic insulators on the surface of an STI have been proposed[17] to gap the surface state and confine the surface electrons to ungapped regions, but this has not been demonstrated experimentally. Recently, another method of opening a bandgap was predicted[21-23] and demonstrated[24, 25]; ARPES showed that ultrathin $Bi_2Se_3$ exhibits a bandgap due to tunnel coupling of top and bottom surfaces[24], and transport experiments revealed that few-layer $Bi_2Se_3$ is a conventional insulator[25]. Here we exploit this effect to create gate-tunable ultrathin $Bi_2Se_3$ barriers to $Bi_2Se_3$ quantum dots. When tunnel barriers are created by gating the electrodes, transport spectroscopy shows Coulomb blockade diamonds with >5 meV charging energy, and evidence of tunneling into excited single-particle states.

The preparation of bulk $Bi_2Se_3$ crystal starting material and its typical carrier densities were described in Reference [26]. $Bi_2Se_3$ thin films were mechanically exfoliated on substrate of 300nm $SiO_2$ with highly doped n-type Si back gate using a "Scotch tape" method typically used for graphene[27]. We often found thin and narrow $Bi_2Se_3$ ribbons cleaved naturally; Figure 1a) shows such a ribbon with a width of 200 nm and a thickness of 7 nm. To fabricate quantum dot devices, electron-beam resist (Microchem Corp. PMMA A4 spun at 5000 rpm) was applied, and electron beam lithography used to define the electrode regions. The channel length $L$ of the shortest devices was ~200nm. After developing to remove the resist in the area of the electrodes, we performed $N_2$ plasma etching at a power of 20 Watts to controllably thin the $Bi_2Se_3$ in the electrode region by a thickness of 3-5 quintuple layers (QLs) before thermal evaporation of Cr/Au (2nm/28nm). Because of the resist undercut typical of electron beam lithography, the region of the $Bi_2Se_3$ film exposed to etching is slightly larger than the deposited metal electrode, creating a narrow region of ultrathin $Bi_2Se_3$ between the electrode and the un-etched $Bi_2Se_3$. The use of a thin single layer of resist minimizes the undercut and ensures narrow barrier regions. We estimate the barrier regions not covered by source and drain have thickness 2-4 QLs and length of order 10 nm. Figure 1d) shows an atomic force microscope (AFM) image of another 100 nm wide and 12 nm thick exfoliated $Bi_2Se_3$ nanoribbon, in which the exposed region after performing e-beam lithography and developing resist was etched to a thickness of 2~4 QLs, but no metal was deposited, allowing AFM imaging of the etched structure.

After etching and source/drain electrode metal deposition, lift-off was done in acetone for 2 hours. All the electrical transport measurements were done in a Quantum Design PPMS 6000 cryostat at a base temperature $T$ = 1.8 K. We note that brief $N_2$ plasma etching of thicker $Bi_2Se_3$ devices immediately before metal electrode deposition is also useful in creating Ohmic contacts between metal electrodes and $Bi_2Se_3$ thin films possibly through removal of surface contaminations or highly disordered surface layers.

Figure 1b) shows an optical image of a completed device. The data reported in Figures 2 and 3 were measured in the short device [dimensions 200 nm ($L$) x 200 nm ($W$) x 7 nm ($t$)] circled in Figure 1b). Figure 1c) shows a schematic of our $Bi_2Se_3$ quantum dot devices. The $Bi_2Se_3$ quantum dot is connected to source and drain electrodes via short and ultrathin $Bi_2Se_3$ films. These ultrathin films are tunable with gate voltage (on and off) and act as tunnel barriers between electrodes and the quantum dot.

Figure 2a shows the dependence of the differential conductance $G$ = d$I$/d$V$ on gate voltage $V_g$ at zero source-drain bias $V$ = 0 for the device in Fig. 1b. A clear off state (zero current) is observed at large negative gate voltages ($V_g$ < -36 V) and an on state (continuous finite current) at when gate voltage is tuned more positive ($V_g$ > -24 V). We believe the origin of the on-off behavior is in the ultrathin barriers since similar metal contacts to thicker $Bi_2Se_3$ are observed to be Ohmic, and 7 nm thick $Bi_2Se_3$ without narrow barriers was not observed to have a hybridized surface state gap[25]. Figure 2b) shows a detail of $G(V_g)$ in the region -36 V < $V_g$ < -24 V.  Quasi-periodic narrow conductance peaks are observed, with no measurable conductance between peaks, reminiscent of Coulomb blockade.

Figure 3 shows transport spectroscopy, i.e. a two-dimensional plot of $G(V,V_g)$, of the gate-voltage region -24.5 V < $V_g$ < -20 V for the device in Fig. 1b.  Diamond-shaped regions of low conductance indicate Coulomb blockade.  The diamonds are fairly regular, and separated by peaks of finite conductance, indicating transport is dominated by a single Coulomb island.  From the average height $\Delta V$ of the Coulomb diamonds, we deduce the charging energy $E_C = e^2/2C_\Sigma$ = 8 meV where $C_\Sigma$ is the total capacitance of the Coulomb island corresponding to $C_\Sigma$ = 20aF. This capacitance agrees well with that estimated from the classical capacitance of a disk $C = 4\kappa\varepsilon_0 \sqrt{A/\pi}$ = 20 aF, where $\varepsilon_0$ is the vacuum permittivity, $\kappa$ = 2.5 is the average dielectric constant for vacuum and SiO$_2$, and $A = L \times W$ = (200 x 200) nm$^2$ is the area of the dot.  This indicates that the 200 nm x 200 nm x 7 nm Bi$_2$Se$_3$ island likely forms a single quantum dot.  The average peak spacing in gate voltage is $\Delta V_g$ = 0.33 V, corresponding to a gate capacitance $C_g$ = 0.5 aF. The bulk of the capacitance is to the leads; the similar slopes of the sides of the Coulomb diamonds indicate roughly similar capacitance to source and drain $C_s \approx C_d \approx$ 10 aF.

Additional enhanced conductance lines parallel to the edge of the Coulomb diamonds were observed outside the Coulomb diamonds and marked by arrows in Figure 3. These lines indicate co-tunneling through excited states of the quantum dot. The lowest excitation energy we observed corresponds to the first excited state energy $\Delta$ ~ 1 meV.  Assuming that the energy quantization occurs from the surface states of Bi$_2$Se$_3$, we find that single-particle energy level spacing $\Delta(N) = \hbar v_F \sqrt{\pi/(NA)}$ [28], where $v_F$ ~ 5 x 10$^5$ m/s is the Fermi velocity[29,30], $A$ is the area of quantum dot and $N$ is the number of electrons in the dot. From $A$ ~ 0.04

μm$^2$ and Δ ~ 1 meV, we estimate the number of electrons in the dot is on the order of 10 (considering that $v_F$ increases with n-type doping, the estimated Δ(*N*) should be considered as a lower bound).

Figure 4 shows transport spectroscopy on a second Bi$_2$Se$_3$ quantum dot device with dimensions 140nm (*W*) x 200nm (*L*) x 6nm (*t*). This device shows similar charging energy and is similar in size to the first device, indicating it also is likely a single Coulomb island. Additional features in transport spectroscopy appear more clearly in this second device as sharp lines running parallel to the edges of the Coulomb diamonds, again indicative of co-tunneling through excited states. Assuming that the excitations are electronic, we again obtain a rough estimate of *N* ~ 10. The number *N* for both devices is surprisingly small considering that in Figure 3 co-tunneling is observed in several diamonds differing in charge by ~5. The small *N* would indicate that both quantum dots have been fortuitously tuned very close to charge neutrality. It is also possible that the co-tunneling reflects some other excitation of the system, e.g. a discrete vibrational mode. More study, for example at lower temperature and in magnetic field, is needed to understand the features of the transport spectroscopy.

In conclusion, we have fabricated topological insulator quantum dots with tunable barriers based on ultrathin Bi$_2$Se$_3$ films. Clear Coulomb blockade was observed with additional features implying excited states. From semiclassical theory, we deduce capacitances of the quantum dot to back gate and electrodes. This study is the essential first step toward topological insulator quantum dot research, and opens up the possibility of studying the quantized modes of the Dirac electronic surface state of topological insulators, including their degeneracy, *g*-factor, level-spacing statistics[31], orbital magnetic moments[32], etc. If topological quantum dots may be connected to suitable topological superconducting leads, the Majorana bound states may be created and studied[16].

Acknowledgements.

We acknowledge support from the UMD-NSF-MRSEC grant DMR-05-20471 and NSF grant DMR-11-05224. NPB acknowledges support from CNAM.

Fig. 1

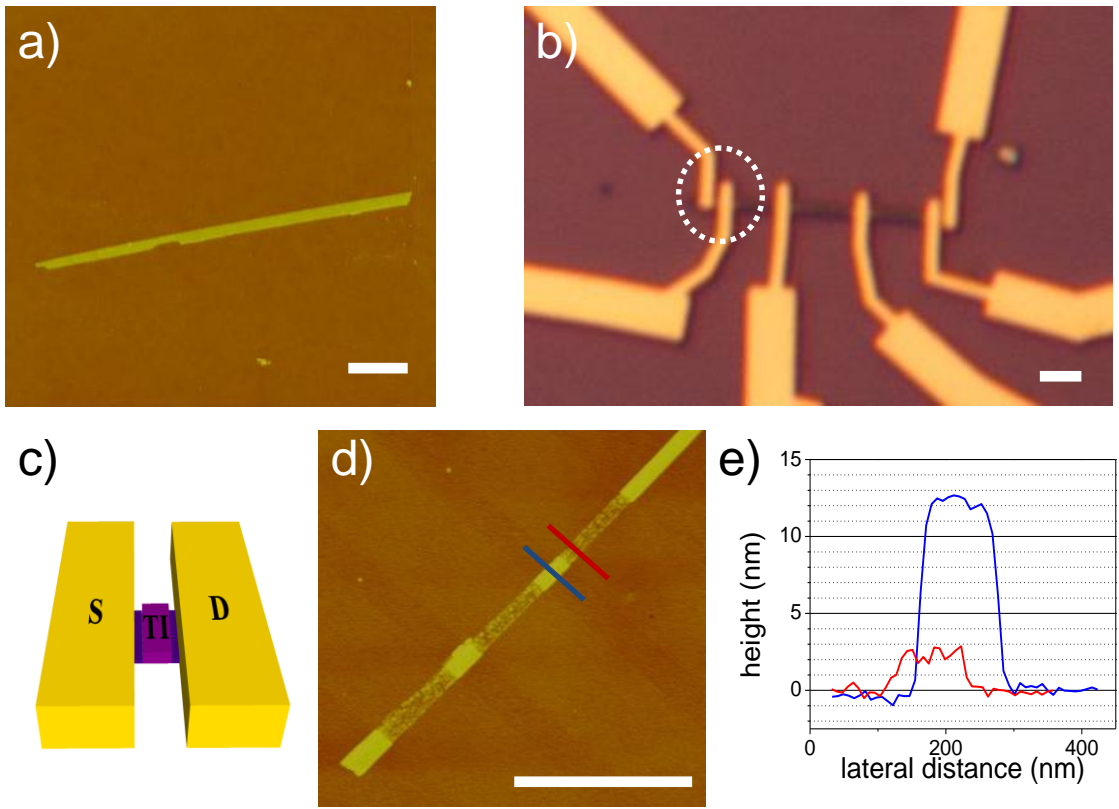

Figure 1.

a) Atomic force micrograph of 7 nm thick, mechanically exfoliated $Bi_2Se_3$ nanoribbon on highly doped $SiO_2$(300nm)/Si. b) optical micrograph of completed device with Cr/Au(2nm/28nm). Dashed circle shows the device used in this study.  c) schematic of $Bi_2Se_3$ quantum dot device. The $Bi_2Se_3$ quantum dot of dimensions 200nm x 200nm x 7nm at the center is connected to source and drain electrodes via short and ultrathin $Bi_2Se_3$ films. d) Atomic force micrograph of 12 nm thick nanoribbon after etching with PMMA mask. e) Line traces of topographic data from (d) along blue line (unetched area) and red line (etched area).  Scale bars correspond to 1μm.

Fig. 2

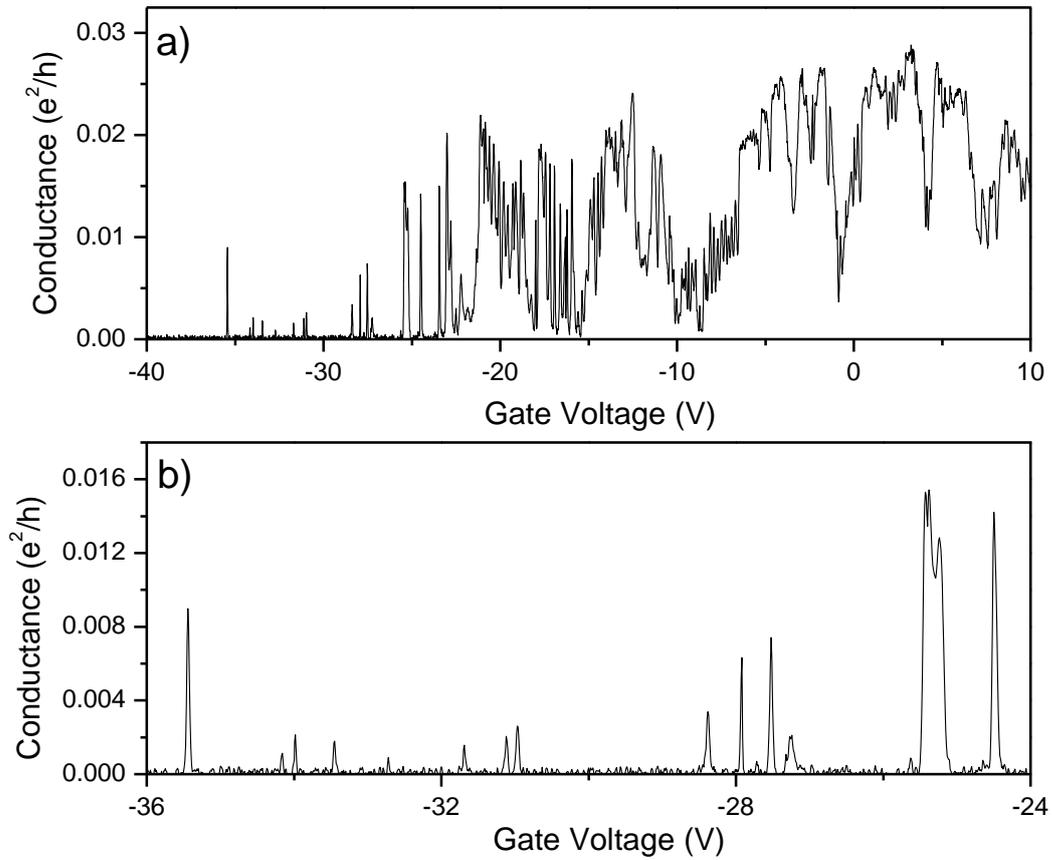

Figure 2.

Gate voltage dependence of differential conductance ($G = dI/dV$) at -40 V < $V_g$ < 10 V (a) and <-36 V < $V_g$ < -24 V (b). In b), quasi-periodic narrow conductance peaks are observed with total suppression of conductance between peaks, indicating Coulomb blockade.

Fig. 3

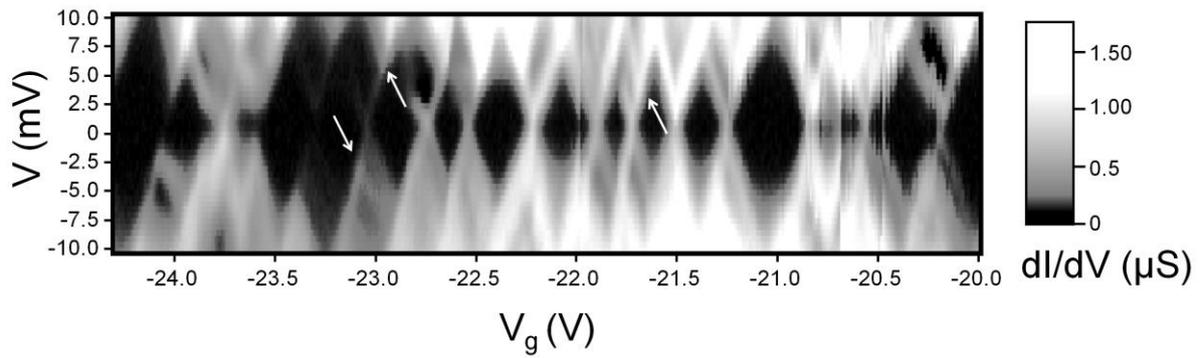

Figure 3.

Two-dimensional plot of conductance *G* as a function of source-drain voltage *V* and gate voltage $V_g$ for -24.5 V < $V_g$ < -20 V. Diamond-shaped regions of low conductance indicate Coulomb blockade. Arrows indicate additional features suggesting excited states of quantum dot energy levels.

Fig. 4

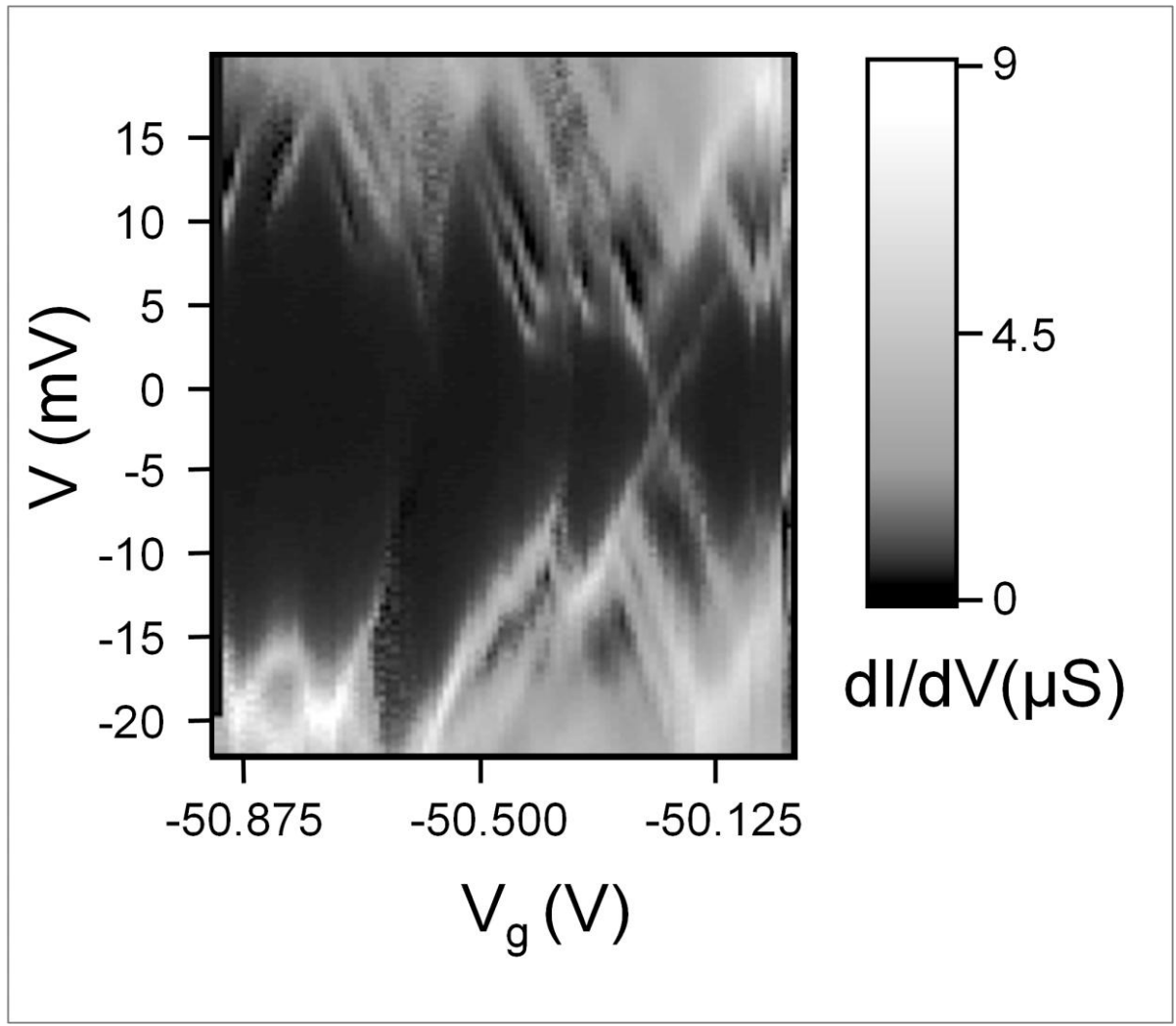

Figure 4.

Two-dimensional plot of conductance *G* as a function of source-drain voltage *V* and gate voltage $V_g$ from an additional $Bi_2Se_3$ quantum dot device showing multiple features (parallel lines outside Coulomb diamonds) suggestive of excited states.